# PREPRINT[1]: Key Lessons from Tailoring Agile Methods for Large-Scale Software Development

*Abstract: We describe advice derived from one of the largest development programs in Norway, where twelve Scrum teams combined agile practices with traditional project management. The Perform program delivered 12 releases over a four-year period, and finished on budget and on time. In this article, we summarize 12 key lessons on five crucial topics, relevant to other large development projects seeking to combine Scrum with traditional project management.*

**By Torgeir Dingsøyr, Tore Dybå, Mette Gjertsen, Anette Odgaard Jacobsen, Tor-Erik Mathisen, Jan Ole Nordfjord, Kjetil Røe, Kjetil Strand**

In the past years, we have seen a major change in how software is developed with the emergence of agile development methods [1]. These methods were believed to best suit small development teams that make software which is not life-critical [2]. However, with the popularity of agile methods, many have started using the methods also in large projects.

Large projects pose great risk and are often associated with cost overruns, late completions and outright project failures [3]. The perils of large-scale development is illustrated by a number of examples such as HealthCare.gov in the US [4]. To ensure successful projects, practitioners using agile methods ask questions like "How do you scale up a large project over many months or even years"[5], and "agile in the large" has been voted the "top burning research question" [6].

Frameworks for managing large agile development projects have started to appear, such as the Scaled Agile Framework [7] and Large-Scale Scrum [8]. However, there are few studies of these frameworks, and the frameworks primarily describe product development, while many organizations choose to establish projects or programs for developing new systems. Projects are different as they are limited in time, will involve setting up a project organization and usually have project participants who need to learn a new domain.

In this article, we describe 12 key lessons from one of the largest development programs in Norway, which provides an example of how twelve Scrum [9] teams combined agile practices with traditional project management [9]. The Perform program (see description in box) delivered a new pension solution after a public reform to the Norwegian Public Service Pension Fund ("the Pension Fund"). The program delivered 12 releases over a four-year period, and finished on budget and on time. In this article, we summarize key advice, which we think is relevant to other large development projects seeking to combine Scrum with traditional project management:

---

[1] (C) IEEE, accepted for publication in *IEEE IT Professional*.

**Backlog management**

The most prominent artifact in agile development is the 'product backlog', which depicts a prioritized queue of high-level 'epics' and 'user stories'. The product backlog is a representation of the scope of an agile software development initiative.

*Lesson1. The product backlog process*: The overall analysis of needs and solution description were made jointly in what was called 'the product backlog process' (see Figure 2). To get the right level of detail in this process, the program management opted for what was 'just enough' in a process of *rolling wave planning*.

At the top level, Epics defined the scope of the Perform program. The Epics were prioritized by importance and were roughly estimated using planning poker [10] so they had a relative size to each other. Through the product backlog process for each release the Epics were broken down into user stories that formed the product backlog items. All product backlog items were prioritized in the order the Product Owners thought they should be performed. However, these priorities could change during the construction process. 'Just enough' in this context, was to detail user stories in the product backlog for two iterations ahead, providing a solution description and an estimate for each user story. To ensure development of high-priority user stories assigned to the right teams at the right time, we recommend rolling wave planning of the product backlog.

*Lesson2. A common backlog*: Initially, three main vendors were responsible for their own subprojects, with three separate product backlogs. This was suboptimal for several reasons:

- It was complicated to move user stories from a vendor/product backlog to another
- It was difficult to prioritize across the backlogs and ensure that the program as a whole was constantly working on the highest priority tasks

The solution to this was to organize the entire program scope in the same product backlog with epics and user stories. From this common product backlog user stories were then distributed to vendors, as they were described more in detail. This enabled that several vendors could work in parallel with the same epic.

A common priority regime across subsystems and vendors ensured that the program constantly worked on the highest priority tasks. It gave a better overview of dependencies between stories, and enabled a more efficient development path. It was easier to communicate and coordinate. Our experience confirms the advice given in most agile methods on having one common backlog.

**Solution descriptions**

From the start, the program planned to have solution descriptions ready for a whole release before starting construction (See Figure 2), in order to ensure that it was high priority user stories that were given to development teams and also to make sure that the description of user stories were of high quality. It was very important that the development teams did not get low-priority items to work on, or that insufficient understanding of user stories led to stories being assigned to the wrong team.

As the program progressed, participants in the Business project (see project organization in box) experienced an increasing workload, focusing on analysis of needs and solution description of coming releases, supporting the development teams with domain knowledge on user stories in the current release, and also clarifying issues regarding releases that were in production. A second challenge was that due to all the learning from feedback and changes in rules and legislation, user stories changed priority before being implemented. This led to resources spent on describing solutions that were not developed. These challenges made the program do two changes in operation:

*Lesson3. Continuous solution description:* User stories were described during the work on one release. Also, the analysis of needs and solution descriptions (Figure 2) phases were merged. This led to efficient use of resources in solution description, as it was only user stories that were going to be implemented that were described, and also the people making the descriptions knew the knowledge of the construction team and could make more concise descriptions. It was easier to predict dependencies between user stories dye to the short time period from solution description to actual implementation.

*Lesson4. Varying level of details:* The teams describing details in varying details up-front. This was due to the nature of work tasks and to working culture in the companies. One of the provider companies wanted more specification up-front in order to reduce rework, while the other focused on more open specifications but continuous collaboration to resolve details. We recommend being open to specifying work to suit needs of the construction teams.

**Coordinating teams**

In large-scale projects where work is divided between many development teams, coordination of the teams is crucial. Early agile methods advised one forum for managing dependencies between the teams, such as the Scrum of Scrums, while Large-Scale Scrum suggests to give the teams the responsibility for coordination and recommend to "just talk". One of the key lessons from the Perform program was how the teams were coordinated, not only using a forum for managing dependencies but through additional roles and additional arenas:

*Lesson 5. Extra roles:* Extra roles were set at the start of the program and implemented in all development teams: Every team had a technical architect responsible for technical design, a functional architect responsible for solution descriptions, a test responsible and a mixture of senior and junior developers (Table 1). This matrix organization (Figure 1) where team members worked also in cross-team projects (Architecture and Test) had several advantages, including saving time by communicating orally, avoiding handovers between subprojects and establishing a feeling of "working on this together". We recommend extra roles to establish coordination between the teams, which in our experience led to development quality, commitment and efficient knowledge sharing.

*Lesson 6. Extra arenas:* To deliver the highest priority user stories and epics early, and to meet external government set milestones, it was necessary to utilize the entire program on the same main components and functionality. The team- and within-iteration dependencies grew. To keep up the speed of the delivery it was important to increase coordination and communication cross teams and vendors. The number of arenas for coordination was much larger than in early advice for agile development. There were daily meetings in the development teams, Scrum of Scrum meetings within each vendor, and a "Metascrum" forum for project managers and subproject managers. In addition, there were arenas for coordination, learning and standardization within the projects Architecture*,* Business and Test. Retrospectives [11] were also held within these groups. There were also a number of informal arenas such as open space meetings and experience sharing fora.

When scaling agile development our experience is that you need additional roles and arenas to ensure efficient coordination between the teams. We recommend a matrix structure to ensure that important concerns are addressed, and to avoid the handover needed if these themes were handled outside of the development teams.

**Quality first**

While Large-Scale Scrum emphasizes agreeing on a "definition of done", Perform took a more formal approach to quality assurance to ensure that the new solution had the right functionality, reliability, user friendliness, performance and maintainability. The strategy was to automate as much of the testing as possible. To handle the scale of the program, we emphasize three changes to standard agile development practices:

*Lesson 7. Test project:* Testing was organized as a separate project with resources from all development teams in addition to a team-external project manager, testers who mainly worked with preparing the approval process described below, and test managers for each of the three development subprojects. This project defined definitions of "done" and acceptance criteria in cooperation with the Business, Development and Architecture projects. At team level, there was one person responsible for making sure that testing took place, but work was divided between all team members. Some teams followed test-driven development, the product was automatically regression-tested every night.

*Lesson 8. Approval process for releases:* A new release went through an approval process in the program before being transferred to acceptance testing, which was conducted by IT operations. This extra process was needed because it is often difficult to verify longer value chains at the last iteration, and the approval process puts a larger emphasis on non-functional requirements such as operability, robustness and performance. To ensure that pensions were calculated correctly, four special test tools were developed to regression test the new solution and compare results with known correct results. These tools included simulated changes in 20,000 pensions, calculation of rights for 8,000 users and regression testing on 250,000 postings for a new solution for settlement. The testers responsible for the approval process did not work in the development teams.

For large programmes, the three lessons above have served to ensure high quality, we recommend that such programs consider these practices.

**Continuous improvement**

Continuous improvement is emphasized in agile methods, mainly through conducting retrospectives at the end of iterations and after releases. An important condition for implementing continuous improvement is support and mandate from enterprise management. This was fully in place in Perform. The program was trusted to implement proposals for organization, processes and concepts, and to adjust these based on experience.

In Perform, the central program management focused on solving problems at the team level whenever possible. Two examples of improvements during the program were establishment of a separate team responsible for development and test environments (a joint venture with the line organization), and that the warranty period for external suppliers was removed after establishing a common product backlog. Some of the most important facilitators of continuous improvement were:

*Lesson 9. Retrospectives:* All teams were required to conduct retrospectives at the end of each iteration, and the minutes were posted on the program wiki. All minutes were read by the central program management, and this feedback from the teams was used to implement changes, and was also used in weekly risk assessments. A team member stated that "this is the first project I have taken part in where the management have been willing to implement changes". Changes were decided in the program management meetings and the Metascrum forum. Sometimes, extra retrospectives were held, for example after having challenges with getting deliverables accepted in the early processes of the program. An internal evaluation of the program, shows that retrospectives were seen as the main instrument for being proactive in continuous learning.

*Lesson 10. Demos as a learning arena:* Teams were given 10 minutes to demonstrate progress after each iteration, and everyone, both in the program and the Pension Fund organization, were invited. In the beginning of the program, there were episodes of team members blaming others when they failed to demonstrate functionality. The central program management held the teams collectively responsible for the progress on their tasks, and this together with developing the team through retrospectives eliminated this problem. The demos were important in communicating what the teams were working on, and the only arena where everyone would be present. Although demos represented a large cost for the program, this was taken because of the importance as a learning arena.

Succeeding with agile methods for large-scale software development is not a matter of course. The method needs to be adapted to changing needs during a programme lifecycle. While there is much good advice in frameworks such as Scaled Agile Framework and Large-Scale Scrum, we believe the advice above will help other programs seeking to combine agile and traditional methods at scale.

**How this article was written**

This article is based on results of a workshop with all authors: Six key participants from the Perform program and two researchers. We started with an open brainstorm on key learning from the program, and structured these into 11 broad groups. Then, we used planning poker to aid our decision as a group to focus on five of the most important topics for this article. We first gave individual votes on how important the topic would be for others, heard arguments for low and high importance, and then gave a final vote. The researchers facilitated a structured discussion on the most important topics, and this material was integrated with an internal experience report, a book on agile contracting and execution where this program is an example [12], and material from 12 separate focus groups covering topics such as project management, inter-team coordination, knowledge management, requirements engineering and architectural work (published in a separate article [9]). The researchers wrote draft sections, which were commented on and expanded by the participants.

**Key terminology**

*Daily meeting* - a short meeting where team members describe work completed, work to be done and any impediments they see for progress within an iteration.

*Demo* - the development team demonstrates completed functionality in a software product to key stakeholders.

*Iteration* - a period, usually of 1-4 weeks, where a team develops new user stories.

*Matrix organization* - many people were both working in development teams focusing on developing user stories, as well as being assigned to the Architecture, Business or Test project with specific responsibilities

*Metascrum meeting* - forum for project managers of all main program projects.

*Retrospectives* - a meeting for a development team to reflect on how the work method could be improved in future iterations.

*Rolling wave planning* - progressive elaboration of plans, "plan a little, do a little".

*Scrum* - agile development method, focuses on evolving a product through small increments, typically involving work from 1-4 weeks.

*Scrum of Scrum meetings* - forum where participants from several Scrum teams would coordinate.

*User story* - a brief statement about a need that a certain user has for functionality in a software solution.

IN A BOX:

**The Perform Program**

The Perform program is one of the largest IT programs in Norway, with a final budget of around EUR 140 million. The program started January 2008 and lasted until March 2012. 175 people were involved in the program, of which 100 were external consultants from five companies. About 800,000 person hours were used to develop around 300 epics, with a total of about 2,500 user stories. These epics were divided into 12 releases. The whole programme was co-located on the same floor.

The program was managed by a program director who mainly focused on external relations, a program manager focusing on the operations, a controller and four project managers responsible for the projects Business, Development, Architecture and Test:

- *Business* - responsible for analysis of needs through defining and prioritizing epics and user stories in a common product backlog. This project was manned with product owners and a total of 30 employees from the line organization in the department. In addition, functional and technical architects from development teams contributed to this project. The project was led by a Pension Fund project manager.
- *Development* - development was divided into three subprojects, one led by the public department, the Pension Fund (6 teams) with own employees and people from five consulting companies. The two other subprojects were led by Accenture (3 teams) and Sopra Steria (3 teams). These feature teams worked according to Scrum with three-week iterations, delivering on a common demonstration day. The feature teams had roles as listed in Table 1. In addition to the 12 feature teams, the project had an environment team responsible for development and test environments. The project was led by a Pension Fund project manager.
- *Architecture* - responsible for defining the overall architecture in the program and also for detailing user stories in the solution description process. Consisted of a lead architect and technical architects from the feature teams. Suppliers Accenture and Sopra Steria participated on a time & material basis. Sometimes, domain experts from the teams participated in solution description. The project was led by an external project manager.
- *Test* - responsible for testing procedures and for approving deliverables from the development teams. Consisted of the test project manager (external), a test manager and testers who mainly worked with preparing the approval process described below and test resources from development teams.

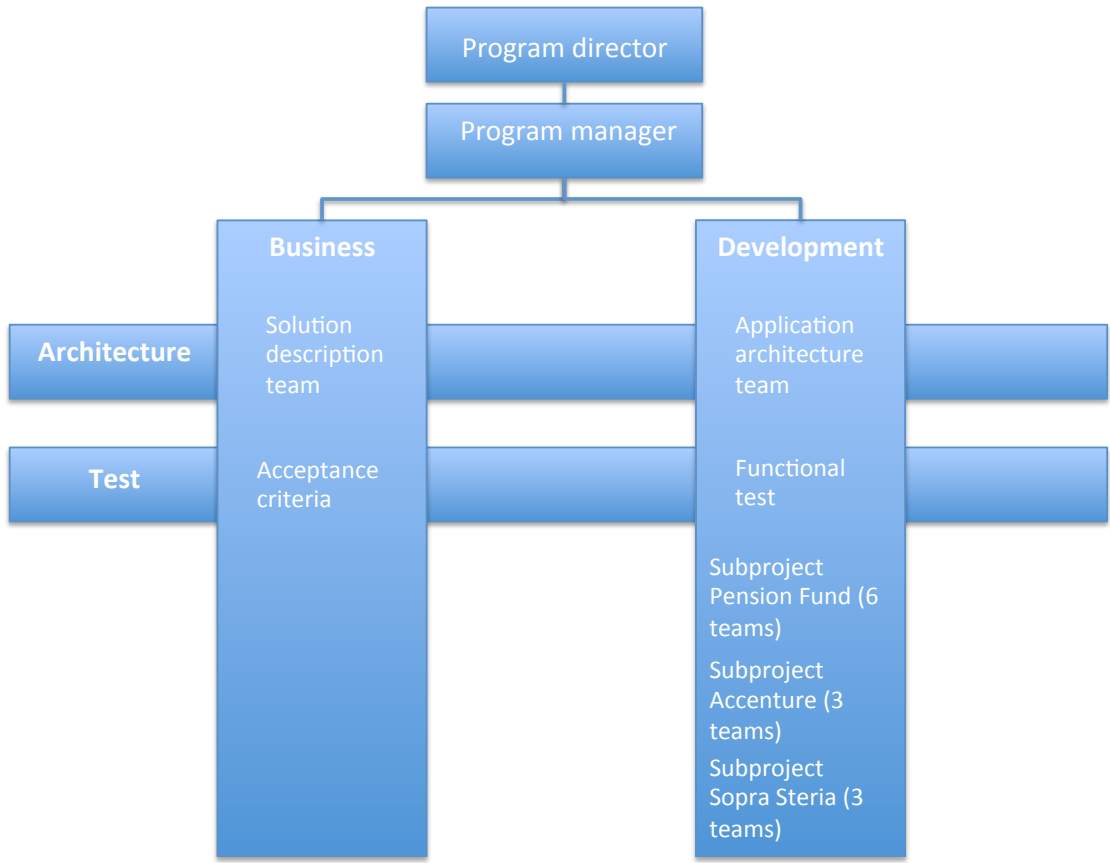

*Figure 1: Program organization, showing the four main projects Business, Development, Architecture and Test. The matrix structure was a key to managing the complexity in the program.*

*Table 1: Roles in the teams in project Development:*

| Role | Description |
|---|---|
| Scrum master | Facilitated daily meetings, iteration planning, demonstration and retrospective. |
| Technical architect | Responsible for technical design, working 50% on this and 50% on development. |
| Functional architect | Responsible for this role was usually allocated 50% to analysis and design, and 50% to development. |
| Test responsible | Made sure that testing was conducted at team level: unit tests, integration tests, system tests and system integration tests. Delivered test criteria to the subproject test. |
| Developers | 4-5 developers were allocated to a team, a mixture of junior and senior developers. |

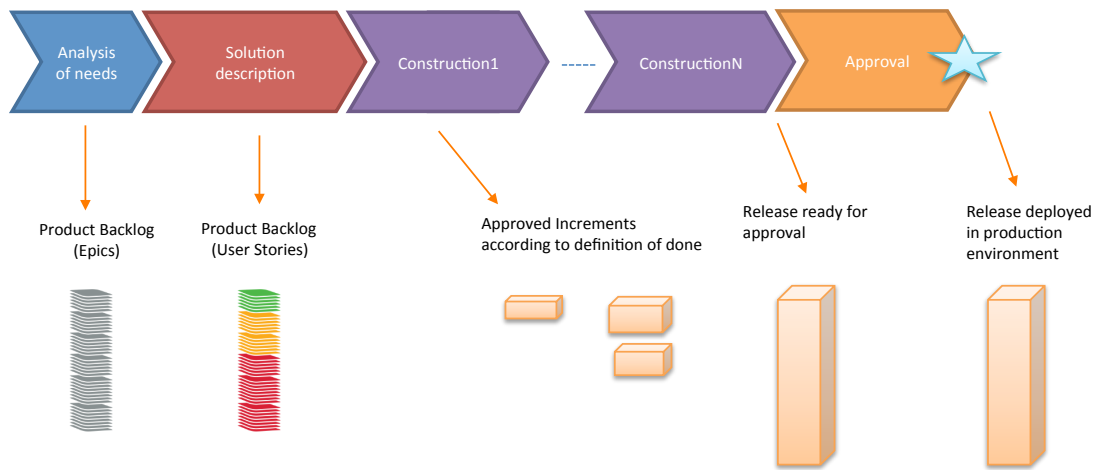

*Figure 2: Initial development process, showing the products of each process. Note that processes were running in parallel for the releases: One release could undergo approval, while another release was under construction and a third in solution description.*

Initially, the development process included the four processes described in Figure 2:

- *Analysis of needs* - this process starts with a walkthrough of target functionality of a release, and identification of epics. The product backlog is prioritized by product owners.
- *Solution description* - epics are divided into smaller user stories, and the user stories are described more in detail, including design and architectural choices. User stories are estimated and assigned to a feature team.

- *Construction* - development and delivery of functionally tested solutions from the product backlog. Three to seven iterations per release.
- *Approval* - a formal functional and non-functional test to verify that the whole release works according to expectations. This includes internal and external interfaces as well at interplay between systems.

In order to keep the schedule of the project, releases were constantly under planning, being constructed and under test.